\documentclass[conference]{IEEEtran}
\IEEEoverridecommandlockouts
\usepackage{cite}
\usepackage{amsmath,amssymb,amsfonts}
\usepackage{algorithmic}
\usepackage{graphicx}
\usepackage{textcomp}
\def\BibTeX{{\rm B\kern-.05em{\sc i\kern-.025em b}\kern-.08em
    T\kern-.1667em\lower.7ex\hbox{E}\kern-.125emX}}
    
\usepackage[dvipsnames]{xcolor}
\usepackage{hyperref}
 
\usepackage{tikz}
\usetikzlibrary{positioning}
\usepackage{enumitem}
\usepackage{subfig}

\begin{document}

%%%%%%%%%%%%%%%%%%%%%%%%%%%%%%%%%%%
%%%% MY PACKAGES %%%%%%%%%%%%
\newcommand{\comment}[1]{}
\newcommand{\tb}[1]{\textcolor{blue}{#1}}
\newcommand{\tr}[1]{\textcolor{red}{#1}}
\newcommand{\tor}[1]{\textcolor{orange}{#1}}
%%%%%%%%%%%%%%%%%%%%%%%%%%%%%%
%%%%%%%%%%%%%%%%%%%%%%%%%%%%%%%%%%%

\title{An FPGA framework for Interferometric Vision-Based Navigation (iVisNav)\\
}

\newcommand{\orcidauthorA}{0000-0002-9191-2130}

\author{\IEEEauthorblockN{Ramchander Rao Bhaskara}
\IEEEauthorblockA{\textit{Dept. of Aerospace Engineering} \\
\textit{Texas A\&M University}\\
College Station, TX 77843, USA \\
bhaskara@tamu.edu}
\and
\IEEEauthorblockN{Kookjin Sung}
\IEEEauthorblockA{\textit{Dept. of Aerospace Engineering} \\
\textit{Texas A\&M University}\\
College Station, TX 77843, USA \\
kookjin.sung@tamu.edu}
\and
\IEEEauthorblockN{Manoranjan Majji}
\IEEEauthorblockA{\textit{Dept. of Aerospace Engineering} \\
\textit{Texas A\&M University}\\
College Station, TX 77843, USA \\
mmajji@tamu.edu}
}
\maketitle

\begin{abstract}
Interferometric Vision-Based Navigation (iVisNav) is a novel optoelectronic sensor for autonomous proximity operations. iVisNav employs laser emitting structured beacons and precisely characterizes six degrees of freedom relative motion rates by measuring changes in the phase of the transmitted laser pulses. iVisNav's embedded package must efficiently process high frequency dynamics for robust sensing and estimation. A new embedded system for least squares-based rate estimation is developed in this paper. The resulting system is capable of interfacing with the photonics and implement the estimation algorithm in a field-programmable gate array. The embedded package is shown to be a hardware/software co-design handling estimation procedure using finite precision arithmetic for high-speed computation. The accuracy of the finite precision FPGA hardware design is compared with the floating-point software evaluation of the algorithm on MATLAB to benchmark its performance and statistical consistency with the error measures. Implementation results demonstrate the utility of FPGA computing capabilities for high-speed proximity navigation using iVisNav.
\end{abstract}


\section{Introduction}

Precise characterization of a vehicle's state is critical to ensure safe navigation operations. Be it spacecraft rendezvous missions or aircraft landing/take-off operations, the state-of-the-art places great emphasis on safe and precise unmanned and autonomous execution \cite{nebot1999sensors}. Technical advancements in microelectronics and embedded systems aid in the autonomy of sensing and control. Reliable control actions demand high-quality information from the sensing devices as well as processing the information in real-time \cite{lopez2013promise}. Hence, rapid enhancements in high-fidelity sensing and computing capabilities advance the real-time execution of autonomous navigation algorithms. 

Traditionally, inertial sensors or inertial measurement units (IMUs) have been used to accomplish the position and attitude sensing for navigation in both autonomous and guided applications \cite{noureldin2012fundamentals,nebot1999sensors}. Although IMUs can capture the dynamics of a fast-moving vehicle, the measurements are typically corrupted by biases, drifts, and noises which accumulate over the rendezvous and lead to considerable errors in pose measurements. Fusing the IMU data with GPS \cite{parkinson1996global} partially addresses this problem by providing another set of measurements (global position) to periodically compensate for the IMU drifts. Close-range rendezvous operations are too critical to entirely depend on the GPS because of bandwidth limitations and ambiguity in resolution for minute positional changes. 

Optoelectronics and machine vision are being embraced at a rapid pace for relative navigation applications \cite{du2005vision, watanabe2004optimal, verras2021vision}. The corresponding vision-based sensing modalities provide rich information context of the surrounding world to the ego-vehicles \cite{tweddle2010computer, christian2013survey}. \comment{Passive vision sensors are not very conducive for applications that request high precision relative pose estimation.} LiDAR sensors in particular, directly provide range measurements in the interest of landing/take-off operations. However, LiDARs are prone to degradation of measurement density with range \cite{gravseth2012vision, majji2011terrain}. Computing range rate from range measurements may lead to large errors and turn out to be unacceptable for precise landing environments. Recent advancements leveraging structured light solutions to realize velocimetry capabilities are found to be robust to most of the issues \cite{junkins2001noncontact, valasek2005vision, wong2016structured, sungspacecraft, sung2022doppler}. 

In addition to sensing, filtering, and optimal state estimation are statutory to effectively utilize the available sensor modalities and thereby achieve mission objectives \cite{crassidis2004optimal}. An online implementation of a filtering approach allows for real-time state estimation, which is critical for autonomous proximity operations \cite{ramchander2021hardware}. Online sensing and filtering algorithms would benefit from execution on dedicated low-cost embedded solutions such as Field Programmable Gate Arrays (FPGAs) \cite{wang2017real}. FPGAs implement customized logic on bare-metal hardware resources and provide infrastructure for processing measurements in real-time.  

Customized FPGA-based embedded solutions deploying hardware-software co-design approaches are highly sought-after in robotics. Modular interfacing with multiple sensing channels, parallel processing of estimation and control schemes at significantly lower footprints is an attractive choice for embedded developers to look away from. Consequently, FPGA accelerated solutions are demonstrated to improve performance in sensor fusion and navigation operations utilizing filtering algorithms \cite{gultekin2013fpga, soh2017scalable, hajdu2017complementary, schaffer2018real,bonato2007fpga,chappell2006exploiting} and digital signal processing \cite{elhossini2006fpga, meher2008fpga}.

In this article, we propose the least squares-based method that leverages FPGA architecture to estimate rate in real-time. We first outline our previous work on sensor design and the estimation process, and present the hardware design next. We demonstrate the functionality of the FPGA framework in simulated results and qualify its performance relative to the legacy software implementation on MATLAB.

%%%%%%%%%%%%%%%%%%%%%%%%%%%%% EXTRA %%%%%%%%%%%%%%%%%%%%%%%%%%%%%%%%%%
\comment{
Imaging sensors assist position estimation from observation through artificial markers or beacons. The methods are very much reliant on the robustness of the software programs that identify feature markers and classify them from false targets. They also consume bandwidth and hence sole dependence on image based navigation might not produce accurate results for the pose. Fusion of vision techniques with inertial sensor measurements for relative pose estimation proves to provide an optimal solution for pose information \cite{verras2021vision}. The multiplicative extended Kalman filter (MEKF) implemented for the fusion, also is reliant on the robustness of the image processing algorithms and imaging techniques to correct for long-term drifts from IMUs. 
}

%%%%%%%%%%%%%%%%%%%%%%%%%%%%%%%%%%%%%%%%%%%%%%%%%%

\section{Related Work}

An analog vision sensor - VisNav is first conceived to overcome the drawbacks of passive vision-based navigation for high precision relative pose estimation \cite{junkins2001noncontact, valasek2005vision}. VisNav uses a set of optical beacons for radiating bursts of frequency-modulated light. Position Sensing Diodes (PSD) on the approaching spacecraft or aircraft sense the modulated optical signals and determine the line of sight toward each beacon. Although the VisNav system supports high-speed optical measurements for position and attitude estimation, it demands a great amount of calibration and installation efforts for the involved analog optoelectronic elements. Alternately, the emergence of high-speed digital cameras replace the PSDs and counters the bandwidth limitations with custom embedded system design \cite{wong2016structured}. The compatibility of monocular camera and LEDs for a variation of VisNav is explored by Wong et al. \cite{wong2016structured}. To achieve the same levels of robustness of VisNav using low-frequency camera measurements, a custom digital hardware design, for filtering and estimation along with sensor data processing is needed.

\begin{figure}[htbp]
\centering
{\includegraphics[width=0.5\textwidth] {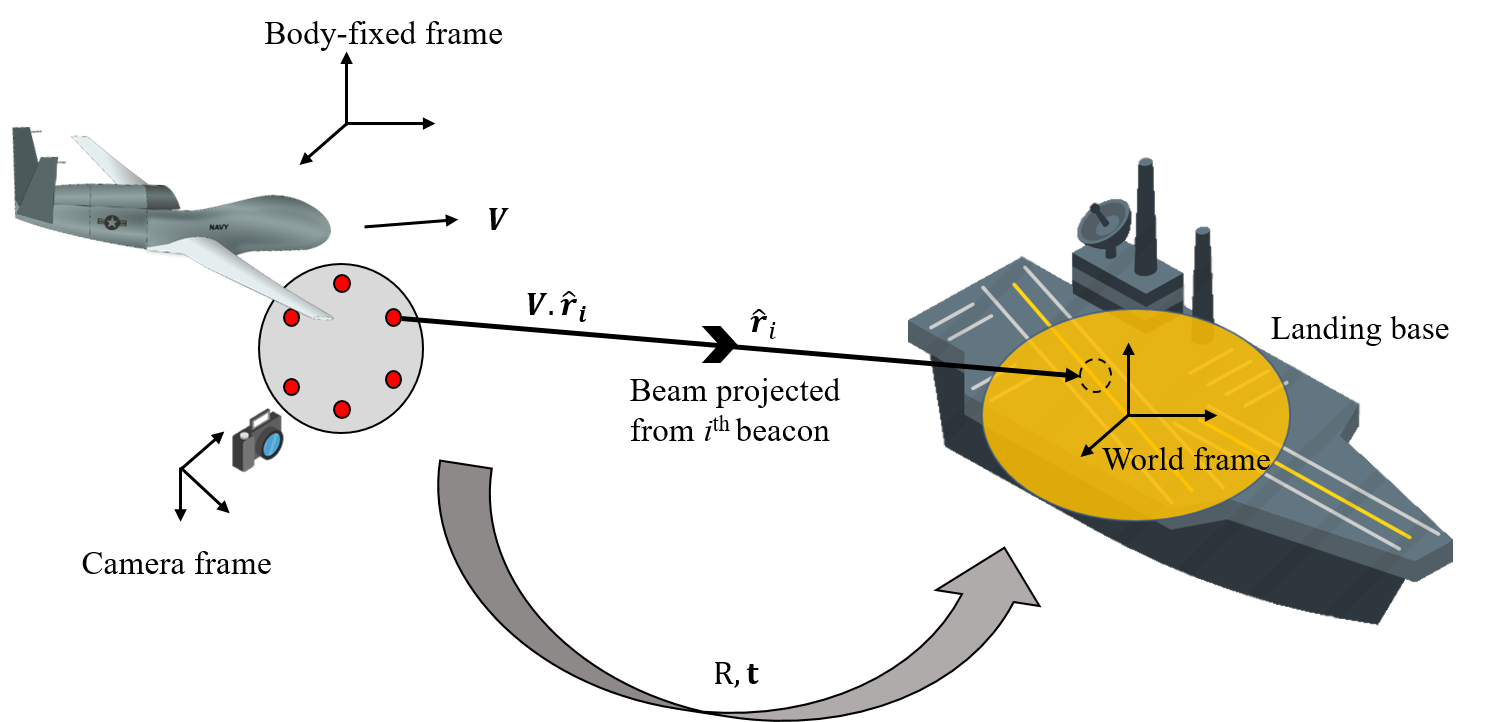}}
\caption{iVisNav sensor system for littoral landing application: Structured beacons project laser pulses onto the landing base. Digital image acquisition locates the projections for relative pose estimation ([$\mathbf{R}$, $\mathbf{t}$]) and the ToF sensors acquire phase shift measurements for relative motion rate estimation. \copyright 2020 IEEE. Modified, with permission, from Sung et al. \cite{sung2020interferometric}.}
\label{fig:ivisnav_sensorSystem}
\end{figure}

Building upon the VisNav system, an Interferometric Vision-based Navigation sensor (iVisNav) \cite{sung2020interferometric} shown in Fig. \ref{fig:ivisnav_sensorSystem}, is proposed for high-frequency velocimetry of a landing base with respect to the sensor system. This is accomplished by illuminating the base with modulated laser light from onboard structured beacons. In one method, Doppler shifts between the illuminated and the reflected light could be captured to estimate the rates of position and angular degrees of freedom \cite{carmer1996laser}. Doppler frequency measurements demand high fidelity and high sampling sensors supported by a high-performance digital computing framework. A simpler and cost-effective approach is to replace the Doppler measurements with phase shift measurements attainable from a Time-of-Flight (ToF) LiDAR. 

Referring to Fig. \ref{fig:ivisnav_sensorSystem}, the ToF LiDAR on-board the aircraft illuminates a landing base with a modulated infrared (IR) laser source and computes the phase shift ($\phi$) of the detected reflection according to Eq. (\ref{eq:iv_phaseShift}). The phase shift measurements translate to the range ($r$) of the moving base from the sensor system as
\begin{equation}\label{eq:iv_phaseShift}
    \phi = 2r \frac{2 \pi}{\lambda} = \frac{4 \pi r f_0}{c}
\end{equation}
\noindent where $\lambda$ is the wavelength of the laser source, $f_0$ is the modulation frequency, and $c$ is the velocity of light ($2.99 \times 10^{8}\; \text{m/s}$).

From consecutive records of phase shift data, a time derivative of the phase is evaluated to derive the relative radial velocity of the aircraft along the projected laser beam. 
\begin{equation}
    \frac{d \phi}{d t} = \frac{4 \pi}{\lambda} \frac{dr}{dt} = \frac{4 \pi v_r}{\lambda}
\end{equation}

This implies that from digitized time-keeping and successive phase shift measurements captured at a high sampling rate, the relative velocity or range rate $v_r$ at $k^{th}$ measurement can be approximated as:
\begin{align} \label{eq:iv_radialVelocity}
    v_r & = \frac{c}{4 \pi f_0} \frac{\Delta \phi}{\Delta t}
\end{align}

The geometric setup of the structured beacon system, bench top prototype, and rate estimation procedures are described in \cite{sung2020interferometric, ramchander2021hardware}. To re-emphasize the algorithmic procedure, least squares-based rate estimation steps are outlined in the next section.

\subsection{iVisNav: Rate estimation}

A set of six or more measurement equations acquired from the iVisNav sensor model assist in the estimation of 6-DOF translational and angular velocity profiles of a rigid body in relative motion. As shown in Eq. (\ref{eq:iv_radialVelocity}), the phase shift measurements (per sampling interval) along the beacon directions $\hat{\mathbf{r}}_i$ are obtained from the ToF LiDAR as
\begin{equation} \label{eq:iv_phaseEquations}
    \Delta \phi_i = \frac{4 \pi (\mathbf{v}_i \cdot \hat{\mathbf{r}}_i)}{c} f_0
\end{equation}

The ToF range measurements are combined with the direction vectors $\hat{\mathbf{r}}_i$ calibrated in the body-fixed frame to obtain range vectors from the beacons to the projections as $\mathbf{r}_i = k_i \hat{\mathbf{r}}_i$ $(i = 1,...,6)$. The relative displacement of the landing base, in the body frame, is denoted by $\mathbf{r}_c$. The scalar $k_i$ is obtained from the ToF sensor's range value measurements. The displacement of beacon projections $\boldsymbol{\rho}_i$ from the origin of the world frame are expressed in terms of the range vectors as 
\begin{equation} \label{eq:iv_r_rho}
    \mathbf{r}_i = \mathbf{r}_c + \boldsymbol{\rho}_i
\end{equation}
\noindent The vectors $\mathbf{r}_i$ and $\mathbf{r}_c$ are coordinatized in the body-fixed frame of reference, while the projection displacements $\boldsymbol{\rho}_i$ are described in the world frame attached to the landing base. 

\noindent The time derivative of Eq. (\ref{eq:iv_r_rho}) is written using the transport theorem \cite{junkins2009analytical} as
\begin{equation}
    \mathbf{v}_i = \mathbf{v}_c + [\boldsymbol{\omega}{\times}]\boldsymbol{\rho}_i
\end{equation}
\noindent where $\boldsymbol{\omega}$ denotes the relative angular velocity vector and $[\boldsymbol{\omega} \times]$ denotes the corresponding skew-symmetric matrix.   

\noindent From Eqs. (\ref{eq:iv_radialVelocity}) and (\ref{eq:iv_phaseEquations}), the  rate measurements from the $i^{th}$ beacon module are re-written as 
\begin{align}
    \mathbf{v}_i \cdot \hat{\mathbf{r}}_i = \frac{\Delta \phi_i}{4 \pi} \lambda &= \mathbf{v}_c \cdot \hat{\mathbf{r}}_i + \hat{\mathbf{r}}_i \cdot [\boldsymbol{\omega}{\times}]\boldsymbol{\rho}_i \notag
    \\
    &= \hat{\mathbf{r}}_i  \cdot \mathbf{v}_c - \hat{\mathbf{r}}_i \cdot [\boldsymbol{\rho}_i{\times}]\boldsymbol{\omega}
\end{align}

\noindent By stacking the system of vectors obtained from each of the six projections, the least squares problem is obtained as 
\begin{gather}
    \frac{\lambda}{4 \pi} \underbrace{\begin{bmatrix}
    \Delta \phi_1 \\ \Delta \phi_2 \\ \vdots \\ \Delta \phi_6
    \end{bmatrix}}_{\tilde{\mathbf{y}}} =
    \underbrace{\begin{bmatrix}
    \hat{\mathbf{r}}_1^T & \vdots & -\hat{\mathbf{r}}_1^T [\boldsymbol{\rho}_1{\times}]
    \\
    \hat{\mathbf{r}}_2^T & \vdots & -\hat{\mathbf{r}}_2^T [\boldsymbol{\rho}_2{\times}]
    \\
    \vdots & \vdots & \vdots 
    \\
    \hat{\mathbf{r}}_6^T & \vdots & -\hat{\mathbf{r}}_6^T [\boldsymbol{\rho}_6{\times}]
    \end{bmatrix}}_{H}
    \begin{bmatrix}
    \mathbf{v}_c \\
    \boldsymbol{\omega}
    \end{bmatrix}
\end{gather}

\noindent The optimal estimate (in the realm of the least squares) for the translational and angular velocity profiles of the center of mass is obtained by the solution to the normal equations as

\begin{equation} \label{eq:iv_leastSquaresAlgo}
    \begin{bmatrix}
    \mathbf{v}_c \\
    \boldsymbol{\omega}
    \end{bmatrix} =  \frac{\lambda}{4 \pi} (H^T W H)^{-1} H^T W {\tilde{\mathbf{y}}}
\end{equation}

\noindent where the $6 \times 6$ symmetric weight matrix $W$ is chosen to be the reciprocal of the measurement error covariance matrix $\Sigma$ such that $W = \Sigma^{-1}$. This choice of $W$ places error proportional emphasis on each of the measurement equations. 

Equation (\ref{eq:iv_leastSquaresAlgo}) is the least squares solution for the estimation of translational and angular rates. The solution demands six phase shift measurements from the structured light setup and also the displacement of the projections $\boldsymbol{\rho}_i$ from utilizing the pose estimation from a low-frequency camera sensor.  

% For completeness we re-visit our previous works in \cite{} 
\section{Hardware Design}

\begin{figure}[htbp]
\centering
{\includegraphics[width=0.5\textwidth] {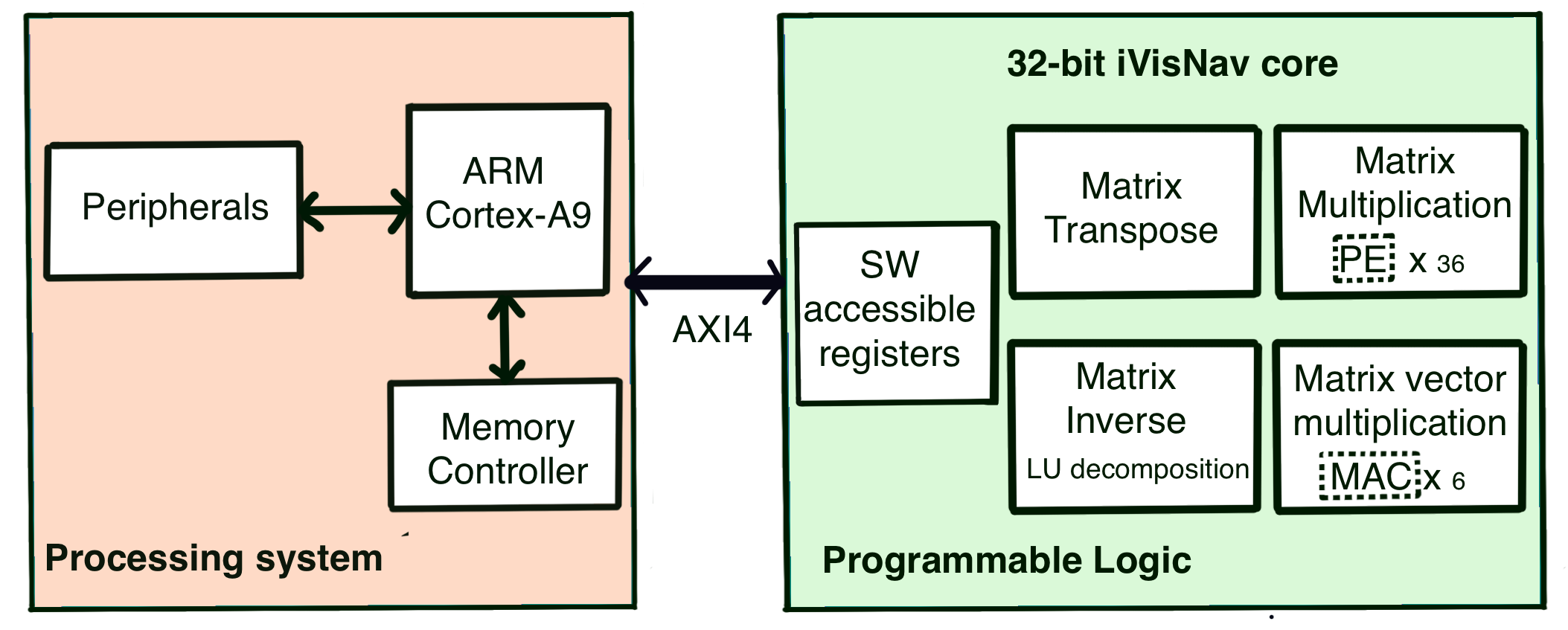}}
\caption{iVisNav HW/SW co-design: The 32-bit integer arithmetic iVisNav core is deployed on programmable logic (PL) to perform the least squares operation. The ARM-based processing system (PS) facilities the streaming of sensor data and estimation results via the AXI4 bus. The PS additionally performs floating-point operations to stream appropriate system and measurement information to the PL.}
\label{fig:ivisnav_core_block}
\end{figure}

The embedded system design for iVisNav estimation framework follows a hardware-software (HW/SW) co-design philosophy, as indicated in Fig. \ref{fig:ivisnav_core_block}. The least squares-based state estimation is realized as a custom intellectual property (IP) core implemented on the programmable logic (PL) of the FPGA. The phase difference measurements and the projection vectors (Eq. (\ref{eq:iv_leastSquaresAlgo})) are generated on the processing system (PS). Simulated measurements on the PS are utilized to validate the least squares implementation. 

\subsection{Development Environment}
HW/SW co-design advantageously combines the traits of development efficiency in software implementation (ARM processor) with high-performance capabilities of the hardware implementation (PL) for the design of FPGA embedded systems \cite{wolf2003decade}. The iVisNav estimation framework is designed to evaluate the repetitive and computationally expensive least squares algorithm on the PL while the application-specific operations are handled by the PS. The application-specific tasks for the iVisNav include measurement pre-processing and data flow control procedures. The embedded system works by efficiently delivering data from PS (via C code running on ARM CPU cores) to the PL (re-configurable FPGA logic, programmed using Verilog hardware description language) for continuous and accelerated evaluation of state estimation sequence. The PS-PL communication is controlled by a state machine (shown in Fig. \ref{fig:iv_statemachine}) and supported by an advanced extensible interface (AXI4) bus protocol utilizing software-accessible registers on the PL. The proposed co-design is implemented on a Xilinx Zynq 7020 FPGA System-on-Chip (SoC) \cite{xilix2016zynq} and programmed using Vivado 2019.1 and Vivado High-Level Synthesis (HLS) tools. Operations on the PL are based on 32-bit fixed-point arithmetic except for matrix inversion operations, which are evaluated using IEEE 754 single-precision floating-point format for extended dynamic range.

% Do we need a Zynq 7020 figure?

\subsection{Architecture: iVisNav Core}
%%%%%
\begin{figure}[htbp]
\centering
{\includegraphics[width=0.45\textwidth] {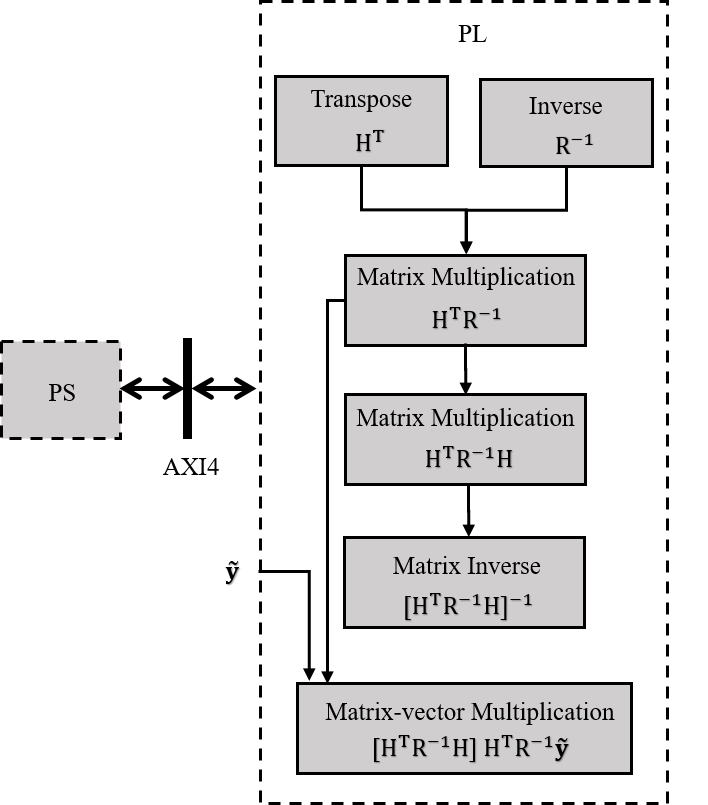}}
\caption{The flow of operations implemented on the iVisNav core for least-squares realization. ARM Processor on the PS communicates and controls the data flow via the AXI4 bus interface.}
\label{fig:iv_ivisnavLogicFlow}
\end{figure}

The pipelined architecture of the iVisNav core is built with a focus on implementing the sequence of linear algebra operations as depicted in Fig. \ref{fig:iv_ivisnavLogicFlow}. Four major modules evaluate the computationally expensive filter operations: matrix transpose, matrix multiplication, matrix inversion, and matrix-vector multiplication. The PS controls the data flow by reading and writing to the software-accessible registers. The PL polls these registers for system and measurement data streams as well as to read/write the status of operations. The implementation overview of the said major modules is as follows: 

\subsubsection{Matrix transpose}

Matrix transpose operation is performed by reshuffling the buffered row-major matrix data stream to output the data in a transposed column-major order. The transpose module utilizes $3\%$ of the lookup tables (LUTs) available on the Zynq 7020 FPGA SoC. 

\subsubsection{Matrix multiplication}

Matrix multiplication is based upon systolic array architecture (SAA) \cite{kung1982systolic}. SAA is a pipelined network arrangement of \textit{Processing Elements} (PEs) in a 2D mesh-like topology \cite{ramchander2021hardware}. The PEs perform multiply and accumulate (MAC) operations on the incoming elements and share this information immediately with the neighboring PEs. SAA avoids repeated memory accesses for matrix elements and thereby is very effective for low-latency matrix multiply operations. The $6 \times 6$ matrix multiplication is an area optimized implementation to meet the stringent resource constraints on the number of digital signal processor (DSP) slices on the Zynq 7020. The high-performance multiplication module occupies $16\%$ of DSPs and $45\%$ of LUTs on board the FPGA.   

\subsubsection{Matrix inverse}
A scalable single-precision floating-point matrix inversion is implemented using LU decomposition algorithm \cite{ruan2017scalable}. Inversion is hardware implemented in stages of: 
\begin{enumerate} [label=(\alph*)]
    \item decomposition of the full-rank matrix $A$, in an iterative manner, to compute a lower triangular matrix $L$, a diagonal matrix $D$, and an upper triangular matrix $U$ as 
\begin{equation}
    A = L \, D \, U
\end{equation}

\item Inversion of the $L,\,D,\,U$ matrices. Special structures enable the computation of their respective inverses with much reduced complexity as shown by Ruan \cite{ruan2017scalable}.

\item Multiplication of $U^{-1}, \, D^{-1}, \, L^{-1}$ to obtain the final output, $A^{-1}$ as 
\begin{equation}
    A^{-1} = U^{-1} D^{-1} L^{-1}
\end{equation}
\end{enumerate}

The inversion module is pipelined at the sub-system level and optimized for high throughput. The complex inversion module is programmed in C and synthesized into register transfer level (RTL) logic using Vivado HLS. The 32-bit inversion module operates using floating-point representation to accommodate a higher dynamic range for internal data representation. To be consistent with the fixed-point implementation of the core, fixed to floating-point and float to fixed-point conversions are performed at the respective input and output ports of the inverse module. Alternative to a floating-point solution, a scaled fixed-point inverse solution might not be able to sustain bit overflows and loss of precision, as commonly observed in higher dimensional matrix operations. This conclusion is based upon failing edge cases in our previous development of a 32-bit fixed-point matrix inversion module using Schur's complement \cite{ramchander2021hardware}. Resource-wise, the inverse module occupies $17\%$ LUTs, $27\%$ DSPs, $9\%$ block RAM, $8\%$ of on-board flip-flops, and $6\%$ of LUT based RAM.

\subsubsection{Matrix vector multiplication}

Analogous to the systolic array architecture, the matrix-vector multiplication utilizes multiply-and-accumulate (MAC) units that operate on the time-aligned input streams of matrix and measurement vector channels. This module takes $11\%$ of LUTs and $12\%$ of DSPs on board the Zynq 7020 FPGA SoC. 

Table \ref{tab:ivisnav_resources} shows the FPGA resource utilization of the iVisNav hardware architecture.

\begin{table}[htbp]
\caption{Implementation requirements for the iVisNav core.}
\begin{center}
\begin{tabular}{|c|c|c|c|}
\hline
{Resource} & {Available} & {Utilization} & {Utilization \%} \\
\hline
LUT & 53200 & 35488 & $66.71$ \\
\hline
LUTRAM & 17400 & 1555 & $8.94$ \\
\hline
FF & 106400 & 14968 & $14.07$ \\
\hline
BRAM & 140 & 12.50 & $8.93$ \\
\hline
DSP & 220 & 120 & $54.55$ \\
\hline
\end{tabular}
\label{tab:ivisnav_resources}
\end{center}
\end{table}

\subsection{State Machine}
\begin{figure}[htbp]
\centering
{\includegraphics[width=0.48\textwidth] {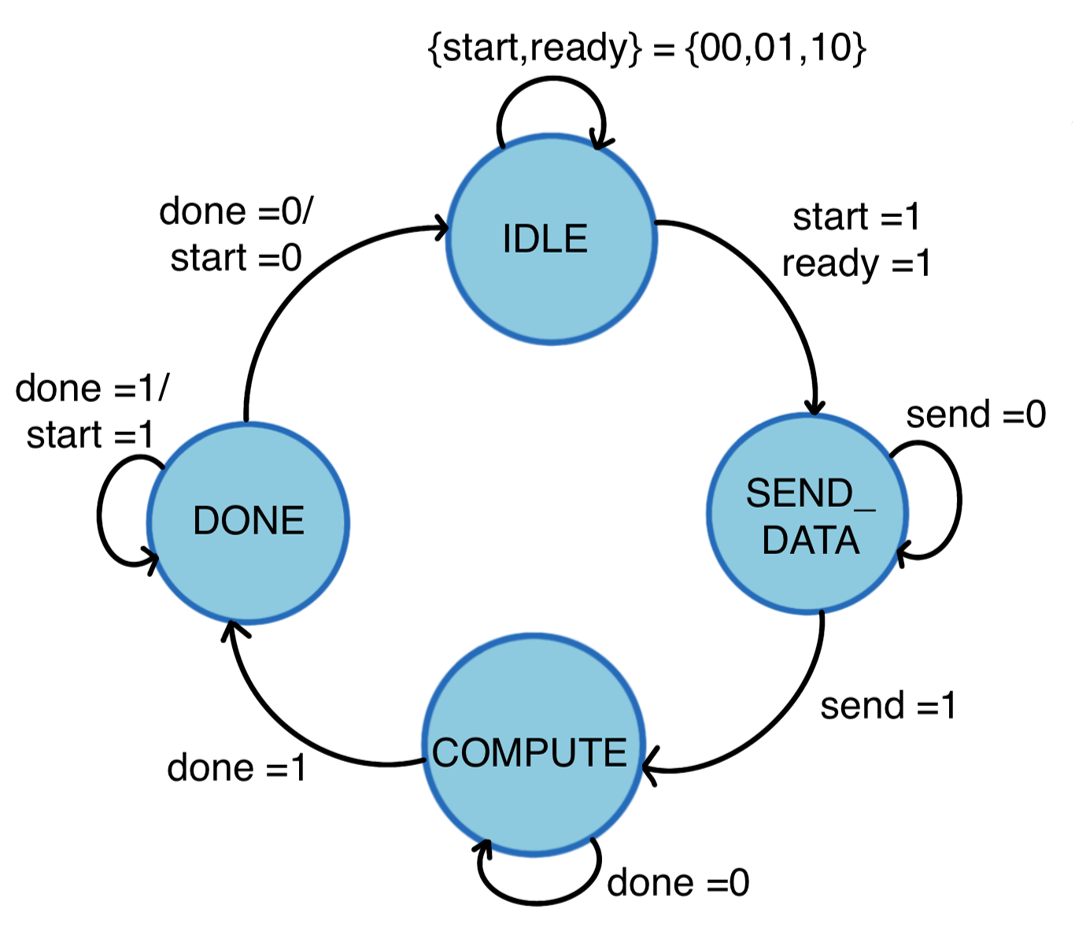}}
\caption{State machine for the iVisNav hardware architecture: The PL remains in \textit{IDLE} state until data is \textit{ready} and the \textit{start} of filtering is asserted. Data is buffered in the \textit{SEND\_DATA} state, and the filtering process takes in the \textit{COMPUTE} state. The PL remains in a \textit{DONE} state until it is asked to restart the cycle of operation.}
\label{fig:iv_statemachine}
\end{figure}

The incoming data stream is buffered on the PL using block memory, and a state machine shown in Fig. \ref{fig:iv_statemachine} controls the data flow on the PL subject to the state of operation. The PL remains \textit{IDLE} until it is \textit{ready}, and a \textit{start} is signalled by the PS. Transfer of data via the software accessible registers takes place until the process of \textit{send}ing is complete. The PL remains in a \textit{COMPUTE} state until the filtering process is \textit{DONE} and the cycle continues.

% Future work: A separate digital signal processing pipeline for the phase measurement acquisition using IQ demodulation is 

\section{Results}

Experimental prototyping of the iVisNav structured light sensor system is demonstrated in \cite{sung2020interferometric, ramchander2021hardware}. Data obtained from the calibrated as well as the simulated sensor platform setup is utilized for validating the proposed hardware-based state estimation. The linear least squares estimation process is studied for analyzing sensitivity to a single axis rotation and translation maneuver of the sensor relative to the projection surface. In this work, we analyze the performance of the fixed-point hardware implementation of the least squares estimation in Eq. (\ref{eq:iv_leastSquaresAlgo}). Double-precision floating-point implementation of the least squares on MATLAB is taken as the golden reference for comparison with the FPGA hardware implementation. 

Sensor calibration process involves configuring the direction vectors $\hat{\mathbf{r}}_i$'s ($i=1,...,6$) in a bench-top experiment \cite{sung2020interferometric}. These values for the said experiment with only axial translation and rotation maneuver are cataloged and shown in Table \ref{tab:iv_directionVectors}. Projection displacements $\boldsymbol{\rho}_i$'s are determined using machine vision (such as in Ref. \cite{sung2020interferometric}). ToF Lidars in the sensor setup deliver the phase shift measurements from which the phase differences are computed. The least squares algorithm shown in Eq. (\ref{eq:iv_leastSquaresAlgo}) is implemented on the acquired data while the projection plate is configured to translate and rotate axially with respect to the beacon setup.

\begin{table}[htbp]
  \caption{Direction vectors corresponding to each of the beacons as configured in the bench top experiment. \copyright 2020 IEEE. Reprinted, with permission, from Sung et al. \cite{sung2020interferometric}.}
\centering
 \begin{tabular}{|c|c|} 
 \hline
 Beacon Index & $\hat{\mathbf{r}}_i$ \\ [0.5ex] 
 \hline
$1$ &  $(0.87264, 0.4977, 0.1367)$ \\
$2$ &  $(0.8927, -0.5082, 0.1304)$ \\
$3$ &  $(-0.0007, -0.9915, 0.1372)$ \\
$4$ &  $(-0.8586, -0.4957, 0.1391)$ \\
$5$ &  $(-0.8168, 0.4957, 0.1412)$ \\
$6$ &  $(0.0001, 0.9999, 0.1249)$ \\
 \hline
 \end{tabular}
  \label{tab:iv_directionVectors}
\end{table}

\begin{figure}[htbp]
\centering
\subfloat[Estimates of translational velocity axial component (where motion is induced).]{\label{fig:iv_hw_v_Est}\includegraphics[width=0.45\textwidth]{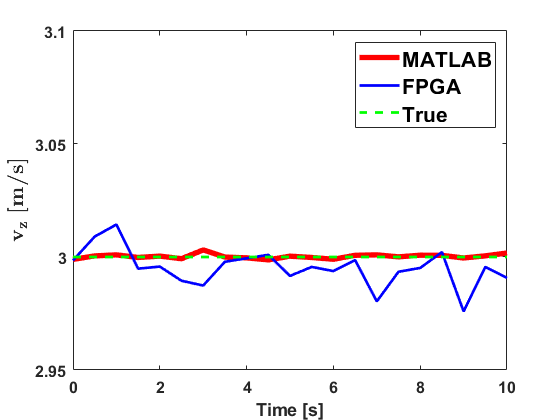}}\qquad
\subfloat[Estimates of angular velocity axial component (where rotation is induced).]{\label{fig:iv_hw_omegaEst}\includegraphics[width=0.45\textwidth]{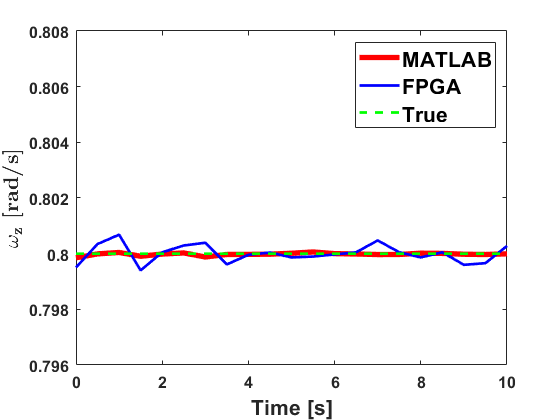}}%
\caption{iVisNav estimation results: The estimates (axial components) from the FPGA implementation (blue) are compared against their true rates (green) respectively. The MATLAB's estimation results (red) are also marked for comparison.}
\label{fig:iv_hw_trueEsts}
\end{figure}

\begin{figure}[htbp]
\centering
\subfloat[Estimates of translational velocity $x$ and $y$ components (no relative motion).]{\label{fig:iv_hw_vxvy}\includegraphics[width=0.45\textwidth]{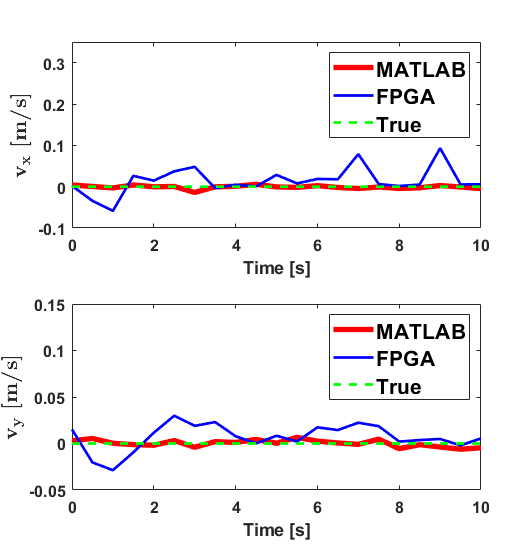}}\qquad
\subfloat[Estimates of angular velocity $x$ and $y$ components (no relative motion).]{\label{fig:iv_hwsw_wx_wy}\includegraphics[width=0.45\textwidth]{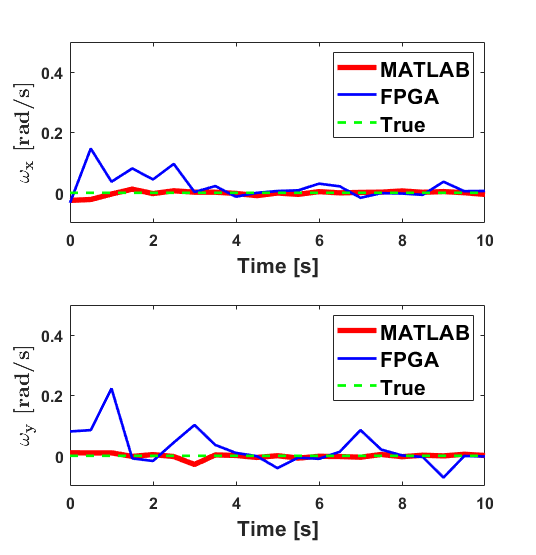}}%
\caption{iVisNav estimation results: The estimates (along the $x$ and $y$ channels) from the FPGA implementation (blue) are compared against their true rates (green) respectively. The MATLAB's estimation results (red) are also marked for comparison.}
\label{fig:iv_hw_trueEsts_xy}
\end{figure}

Hardware implementation of the iVisNav core is validated with the simulated inputs that correspond to system and covariance matrices $H$ and $R$ respectively, as well as the phase shift measurements, $\tilde{\mathbf{y}}$. The results obtained from the hardware implementation are compared with the true rates and are shown in Figs. \ref{fig:iv_hw_trueEsts} and \ref{fig:iv_hw_trueEsts_xy}. The figure also juxtaposes the estimates from the MATLAB's implementation for comparing the hardware implementation accuracy with that of the MATLAB's. We report relatively larger deviations from MATLAB's results along $v_x, v_y, \omega_x, \omega_y$ channels where no motion is induced. The finite precision quantization errors appear to dominate the fractional bit representation while representing near $0$ values. These errors appear to have propagated through the matrix operations, yielding the deviations. Scaling the data prior to finite precision processing, higher number of fractional bits for representation, and floating-point representation are some techniques observed to mitigate this issue and these alternative solutions are being studied to improve accuracy in the edge-cases of filter implementation.
\comment{
The errors in each of the six channels for the estimated translational and angular velocities obtained from the iVisNav core are plotted in Fig. \ref{fig:iv_hw_ErrEsts}. The errors along all the channels are observed to be statistically bound within their $3-\sigma$ covariance limits.}

\begin{figure}[ht]
\centering
\includegraphics[width=0.45\textwidth]{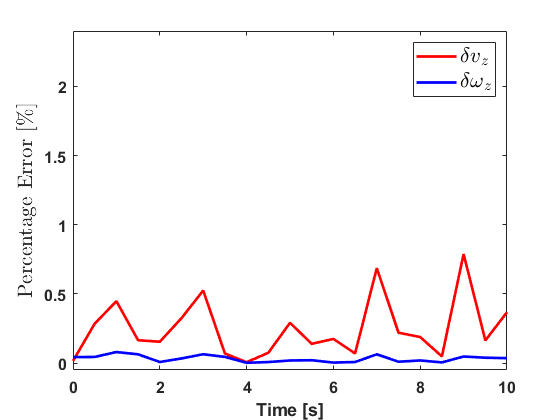}
\caption{Percentage errors in $v_z$ and $\omega_z$ components: MATLAB simulation results are used as reference for relative error computation.}
\label{fig:iv_HWSWPercentages}
\end{figure}

The percentage of relative errors in the channels $v_z$ and $\omega_z$ (where motion is induced) are shown in Fig. \ref{fig:iv_HWSWPercentages}. The percentage error for an estimate $\hat{{x}}$ obtained at the $i^{th}$ instance from the hardware (HW) in contrast with the software (SW) is computed as indicated in Eq. (\ref{eq:iv_errorPercent}). These errors are a relative comparison between the output obtained from MATLAB to that obtained from the iVisNav core's hardware simulation. The errors are below $0.8\%$ which indicates the accuracy that our 32-bit finite precision hardware implementation offers. Although this error performance is not guaranteed across the range of measurements, but scaling the data appropriately to represent it in the 32 available bits is paramount to achieving lower error percentages ($<5\%$). The latency of the least-squares implementation, measured via simulation, is observed to be about $7.1$ microseconds. The results demonstrate that the Register Transfer Level (RTL) design for the FPGA based estimation reliably replicates a software implementation while offering high-speed compute capabilities.    

\begin{equation} \label{eq:iv_errorPercent}
    \% \text{Error} = \frac{|\hat{{x}}_{i,\text{HW}} - \hat{{x}}_{i,\text{SW}}|}{|\hat{{x}}_{i,\text{SW}}|} \times 100
\end{equation}

\section{Conclusion and Future Work}
A finite-precision FPGA framework to estimate the relative rates of a moving projection base with respect to the iVisNav sensor system is developed in this paper. While designed for electro-optical sensors like iVisNav, the framework is applicable for motion rate estimation utilizing other Doppler sensors that use RF or other forms of energy modulation. The embedded system is capable of achieving high data throughput and accuracy requirements for real-time navigation tasks. Simulated phase difference measurements are used to verify the functioning of the iVisNav estimation logic on the FPGA. The estimation results from the 32-bit fixed-point implementation are observed to have a deviation less than {$1\%$} from a double-precision MATLAB implementation of the same method and with a compute latency of $7.1$ $\mu$s. Although this conservative error performance is specific to the presented simulation case and carefully chosen data scaling scheme, the FPGA implementation is observed to be accurate to the quantization bandwidth and forms a basis for optimism to potentially replace floating-point operations. Thereby, the authors conclude that similar approaches can be effectively used for high-speed pipelined frameworks and advanced sensing architectures of the future. 

The framework is designed as a standalone IP core for interfacing with external modules. The state estimation was found to be realizable in a flight system by implementing it on an FPGA combined with a high-speed digitizer. Robust estimation pipelines are also being researched to improve performance and accuracy in next-generation embedded avionics. 

\section*{Acknowledgment}

This work is supported by the Office of Naval Research Grant Number N00014-19-1-2435. The authors acknowledge Dr. Brian Holm-Hansen and Dr. David Gonzales for their support. The authors are also thankful to Peter Arslanian, Daniel Shafer, NAWCAD, Pax-River for their support.

\bibliographystyle{ieeetr}
\bibliography{references.bib}

\begin{thebibliography}{10}

\bibitem{nebot1999sensors}
E.~Nebot, ``Sensors used for autonomous navigation,'' in {\em Advances in
  Intelligent Autonomous Systems}, pp.~135--156, Springer, 1999.

\bibitem{lopez2013promise}
S.~Lopez, T.~Vladimirova, C.~Gonzalez, J.~Resano, D.~Mozos, and A.~Plaza, ``The
  promise of reconfigurable computing for hyperspectral imaging onboard
  systems: A review and trends,'' {\em Proceedings of the IEEE}, vol.~101,
  no.~3, pp.~698--722, 2013.

\bibitem{noureldin2012fundamentals}
A.~Noureldin, T.~B. Karamat, and J.~Georgy, {\em Fundamentals of inertial
  navigation, satellite-based positioning and their integration}.
\newblock Springer Science \& Business Media, 2012.

\bibitem{parkinson1996global}
B.~W. Parkinson, P.~Enge, P.~Axelrad, and J.~J. Spilker~Jr, {\em Global
  positioning system: Theory and applications, Volume II}.
\newblock American Institute of Aeronautics and Astronautics, 1996.

\bibitem{du2005vision}
J.-Y. Du, {\em Vision based navigation system for autonomous proximity
  operations: an experimental and analytical study}.
\newblock PhD thesis, Texas A\&M University, 2005.

\bibitem{watanabe2004optimal}
Y.~Watanabe, E.~Johnson, and A.~Calise, ``Optimal 3-d guidance from a 2-d
  vision sensor,'' in {\em AIAA Guidance, Navigation, and Control Conference
  and Exhibit}, p.~4779, 2004.

\bibitem{verras2021vision}
A.~Verras, R.~T. Eapen, A.~B. Simon, M.~Majji, R.~R. Bhaskara, C.~I. Restrepo,
  and R.~Lovelace, ``Vision and inertial sensor fusion for terrain relative
  navigation,'' in {\em AIAA Scitech 2021 Forum}, p.~0646, 2021.

\bibitem{tweddle2010computer}
B.~E. Tweddle, {\em Computer vision based navigation for spacecraft proximity
  operations}.
\newblock PhD thesis, Massachusetts Institute of Technology, 2010.

\bibitem{christian2013survey}
J.~A. Christian and S.~Cryan, ``A survey of lidar technology and its use in
  spacecraft relative navigation,'' in {\em AIAA Guidance, Navigation, and
  Control (GNC) Conference}, p.~4641, 2013.

\bibitem{gravseth2012vision}
I.~J. Gravseth, R.~Rohrschneider, and J.~Masciarelli, ``Vision navigation
  sensor(vns) results from the storrm mission,'' {\em Advances in the
  Astronautical Sciences}, vol.~144, pp.~223--242, 2012.

\bibitem{majji2011terrain}
M.~Majji, J.~Davis, J.~Doebbler, J.~Junkins, B.~Macomber, M.~Vavrina, and
  J.~Vian, ``Terrain mapping and landing operations using vision based
  navigation systems,'' in {\em AIAA Guidance, Navigation, and Control
  Conference}, p.~6581, 2011.

\bibitem{junkins2001noncontact}
J.~L. Junkins, D.~Hughes, and H.~Schaub, ``Noncontact position and orientation
  measurement system and method,'' July~24 2001.
\newblock US Patent 6,266,142.

\bibitem{valasek2005vision}
J.~Valasek, K.~Gunnam, J.~Kimmett, M.~D. Tandale, J.~L. Junkins, and D.~Hughes,
  ``Vision-based sensor and navigation system for autonomous air refueling,''
  {\em Journal of Guidance, Control, and Dynamics}, vol.~28, no.~5,
  pp.~979--989, 2005.

\bibitem{wong2016structured}
X.~I. Wong and M.~Majji, ``A structured light system for relative navigation
  applications,'' {\em IEEE Sensors Journal}, vol.~16, no.~17, pp.~6662--6679,
  2016.

\bibitem{sungspacecraft}
K.~Sung and M.~Majji, ``Spacecraft proximity navigation using the ivisnav
  sensor system,''

\bibitem{sung2022doppler}
K.~Sung and M.~Majji, ``Doppler measurement of modulated light for high speed
  vehicles,'' {\em Sensors}, vol.~22, no.~4, p.~1444, 2022.

\bibitem{crassidis2004optimal}
J.~L. Crassidis and J.~L. Junkins, {\em Optimal estimation of dynamic systems}.
\newblock Chapman and Hall/CRC, 2004.

\bibitem{ramchander2021hardware}
B.~Ramchander~Rao, ``Hardware implementation of navigation filters for
  automation of dynamical systems,'' Master's thesis, Texas A\&M University,
  2021.

\bibitem{wang2017real}
C.~Wang, E.~D. Burnham-Fay, and J.~D. Ellis, ``Real-time fpga-based kalman
  filter for constant and non-constant velocity periodic error correction,''
  {\em Precision Engineering}, vol.~48, pp.~133--143, 2017.

\bibitem{gultekin2013fpga}
G.~K. Gultekin and A.~Saranli, ``An fpga based high performance optical flow
  hardware design for computer vision applications,'' {\em Microprocessors and
  Microsystems}, vol.~37, no.~3, pp.~270--286, 2013.

\bibitem{soh2017scalable}
J.~Soh, {\em A scalable, portable, FPGA-based implementation of the Unscented
  Kalman Filter}.
\newblock PhD thesis, 2017.

\bibitem{hajdu2017complementary}
S.~Hajdu, S.~T. Brassai, and I.~Szekely, ``Complementary filter based sensor
  fusion on fpga platforms,'' in {\em 2017 International Conference on
  Optimization of Electrical and Electronic Equipment (OPTIM) \& 2017 Intl
  Aegean Conference on Electrical Machines and Power Electronics (ACEMP)},
  pp.~851--856, IEEE, 2017.

\bibitem{schaffer2018real}
L.~Sch{\"a}ffer, Z.~Kincses, and S.~Pletl, ``A real-time pose estimation
  algorithm based on fpga and sensor fusion,'' in {\em 2018 IEEE 16th
  International Symposium on Intelligent Systems and Informatics (SISY)},
  pp.~000149--000154, IEEE, 2018.

\bibitem{bonato2007fpga}
V.~Bonato, R.~Peron, D.~F. Wolf, J.~A. de~Holanda, E.~Marques, and J.~M.
  Cardoso, ``An fpga implementation for a kalman filter with application to
  mobile robotics,'' in {\em 2007 International Symposium on Industrial
  Embedded Systems}, pp.~148--155, IEEE, 2007.

\bibitem{chappell2006exploiting}
S.~Chappell, A.~Macarthur, D.~Preston, D.~Olmstead, B.~Flint, and C.~Sullivan,
  ``Exploiting real-time fpga based adaptive systems technology for real-time
  sensor fusion in next generation automotive safety systems,'' in {\em The IEE
  Seminar on Target Tracking: Algorithms and Applications 2006 (Ref. No.
  2006/11359)}, pp.~61--68, IET, 2006.

\bibitem{elhossini2006fpga}
A.~Elhossini, S.~Areibi, and R.~Dony, ``An fpga implementation of the lms
  adaptive filter for audio processing,'' in {\em 2006 IEEE International
  Conference on Reconfigurable Computing and FPGA's (ReConFig 2006)}, pp.~1--8,
  IEEE, 2006.

\bibitem{meher2008fpga}
P.~K. Meher, S.~Chandrasekaran, and A.~Amira, ``Fpga realization of fir filters
  by efficient and flexible systolization using distributed arithmetic,'' {\em
  IEEE transactions on signal processing}, vol.~56, no.~7, pp.~3009--3017,
  2008.

\bibitem{sung2020interferometric}
K.~Sung, R.~Bhaskara, and M.~Majji, ``Interferometric vision-based navigation
  sensor for autonomous proximity operation,'' in {\em 2020 AIAA/IEEE 39th
  Digital Avionics Systems Conference (DASC)}, pp.~1--7, IEEE, 2020.

\bibitem{carmer1996laser}
D.~C. Carmer and L.~M. Peterson, ``Laser radar in robotics,'' {\em Proceedings
  of the IEEE}, vol.~84, no.~2, pp.~299--320, 1996.

\bibitem{junkins2009analytical}
J.~L. Junkins and H.~Schaub, {\em Analytical mechanics of space systems}.
\newblock American Institute of Aeronautics and Astronautics, 2009.

\bibitem{wolf2003decade}
W.~Wolf, ``A decade of hardware/software codesign,'' {\em Computer}, vol.~36,
  no.~4, pp.~38--43, 2003.

\bibitem{xilix2016zynq}
{Xilinx, Inc}, ``Zynq-7000 all programmable {SoC}: Technical reference manual,
  v1.12.2,'' 2018.
\newblock
  \url{{https://www.xilinx.com/support/documentation/user\_guides/ug585-Zynq-7000-TRM.pdf}},
  Last accessed on 2022-06-24.

\bibitem{kung1982systolic}
H.-T. Kung, ``Why systolic architectures?,'' {\em IEEE computer}, vol.~15,
  no.~1, pp.~37--46, 1982.

\bibitem{ruan2017scalable}
M.~Ruan, ``Scalable floating-point matrix inversion design using vivado high
  level synthesis,'' in {\em Application Notes}, pp.~1--20, Xilinx, 2017.

\end{thebibliography}

\end{document}